\documentclass[preprint,12pt]{elsarticle}
\usepackage{float}
\usepackage{lineno}

\biboptions{sort&compress}

\begin{document}

\begin{frontmatter}

\title{The water system and radon measurement system of Jiangmen Underground Neutrino Observatory}

\renewcommand{\thefootnote}{\fnsymbol{footnote}}
\author{
C.~Guo$^{a}\footnote{Corresponding author. Tel:~+86-01088236256. E-mail address: guocong@ihep.ac.cn (C.~Guo).}$,
Y.P.~Zhang$^{a}$,
J.C.~Liu$^{a}$,
C.G.~Yang$^{a}$,
P.~Zhang$^{a}$
}
\address{
${^a}$Key Laboratory of Particle Astrophysics, Institute of High Energy Physics, Chinese Academy of Science, Beijing, China\\
}

\begin{abstract}
The Jiangmen Underground Neutrino Observatory (JUNO), a 20~ktons multi-purpose underground liquid scintillator detector, was proposed with the determination of the neutrino mass hierarchy as a primary physics goal. Due to low background requirement of the experiment, a multi-veto system ,which consists of a water Cherenkov detector and a top tracker detector, is required. In order to keep the water quality good and remove the radon in the water, a ultra-pure water system, a radon removal system and radon concentration measurement system have been designed. In this paper, the radon removal equipments and its radon removal limit will be presented.
\end{abstract}


\end{frontmatter}


\section{Introduction}
\begin{figure}[htb]
\centering
\includegraphics[width=8cm]{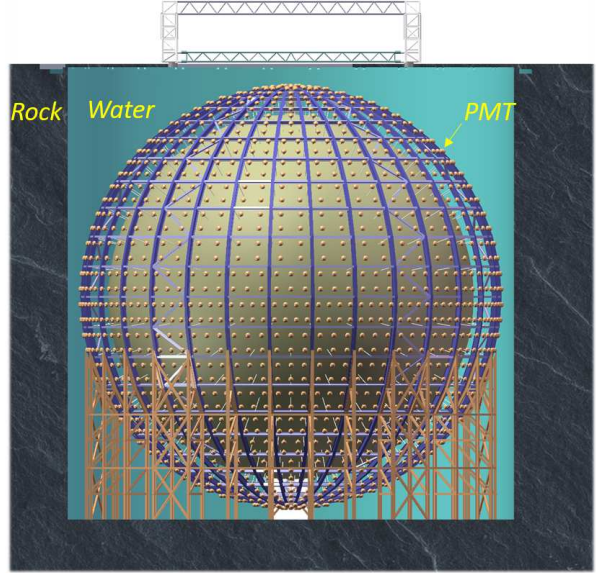}
\caption{The conceptual design of the water Cherenkov detector. }
\label{fig.juno_veto}
\end{figure}

The Jiangmen Underground Neutrino Observatory(JUNO)~\cite{JUNO,JUNO_physics} is a multipurpose neutrino experiment designed to determine neutrino mass hierarchy and precisely measure oscillation parameters by detecting reactor neutrinos from the Yangjiang and Taishan Nuclear Power Plants, with a 20-thousand-tons liquid scintillator(LS) detector of unprecedented 3\% energy resolution (at 1 MeV) at 700-meters deep underground. To suppress radioactivity and cosmogenic background, the outer of the central detector is filled with ultra-pure water as passive shielding for radioactivity from surrounding rock and is equipped with about 2000 microchannel plate photomultiplier tubes( MCP-PMTs, 20~inchs) to form a water Cherenkov veto detector to tag muons. The conceptual design of the water Cherenkov detector is shown in Fig.~\ref{fig.juno_veto}.

Three basic requirements have been put forward for the ultra-pure water system:

(A)There will be different kinds of materials submerged in the water, including stainless steel, Tyvek, PMT glass, cables, etc. The complexity of the underground environment makes it difficult to seal the pool completely and almost impossible to keep the water quality good for a long time. Therefore it is necessary to build a reliable ultra-pure water production, purification and circulation system~\cite{JUNO}.

(B)The temperature stability of the central detector is critical for the entire experiment,thus one of the most important function of the water system is to keep the overall detector temperature stable~\cite{JUNO}.

(C)According to the MC simulation of JUNO experiment requirements, the radon concentration in the water should be less than 0.2~Bq/m$^3$\cite{JUNO_physics}. However, radon can be emanated from the surface of various radium containing substances, including the wall of the water pool, the PMT glass, the stainless steel, etc. Thus the water system should have the function of removing the radon in water.

\section{The ultra-pure water production and circulation system}
\begin{figure}[htb]
\centering
\includegraphics[width=10cm]{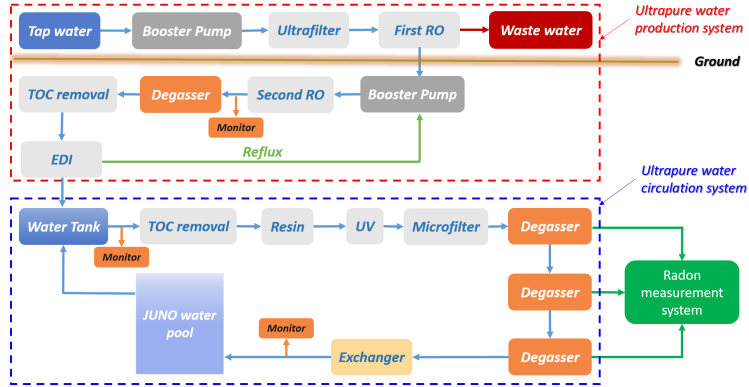}
\caption{The conceptual scheme of the water system.}
\label{fig.watersystem}
\end{figure}

The conceptual scheme of the JUNO water system, which consists of the ultra-pure water production system and the circulation system, is shown in Fig.~\ref{fig.watersystem}. The function of each equipment is as follows:

(A) Ro: Reverse Osmosis(RO) is  a water purification technology, which uses a semipermeable membrane to remove the dissolved ions, molecules as well as the suspended species in water~\cite{RO};

(B) TOC removal: TOC is short for total organic carbon. This device can be used to remove the organic compound in water~\cite{TOC};

(C) EDI: Electrodeionization(EDI) is another kind of  water treatment technology that utilizes electricity and ion exchange membranes to deionize water and separate dissolved ions/impurities from water~\cite{EDI};

(D) Resin: It can be used to remove the dissolved ions;

(E) UV: The attenuation length of water can be affected by the number of bacteria in it. The UV is used to sterilize;

(F) Exchanger: A chiller is included in this part and its function is to control the temperature of the water;

\begin{figure}[htb]
\centering
\includegraphics[width=8cm]{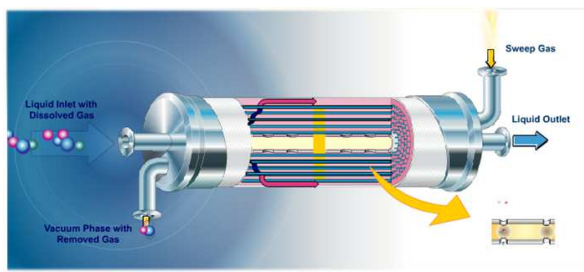}
\caption{The schematic diagram of the membrane.}
\label{fig.degasser}
\end{figure}

(G) Degasser: The main part of degasser is the Liqui-Cel Membrane contactor, which is used for degassing liquid. They are widely used for oxygen and carbon dioxide removal. A membrane contactor can contain thousands of microporous polypropylene hollow fibers knitted into an array that is wound around a center tube. The hollow fibers are arranged with uniform spacing, allowing greater flow capacity and utilization of the total membrane surface area. Because the hollow fiber membrane is hydrophobic, liquids will not penetrate the membrane pores. The schematic diagram of the membrane is shown in Fig~\ref{fig.degasser}.

\begin{figure}[H]
\centering
\includegraphics[width=8cm]{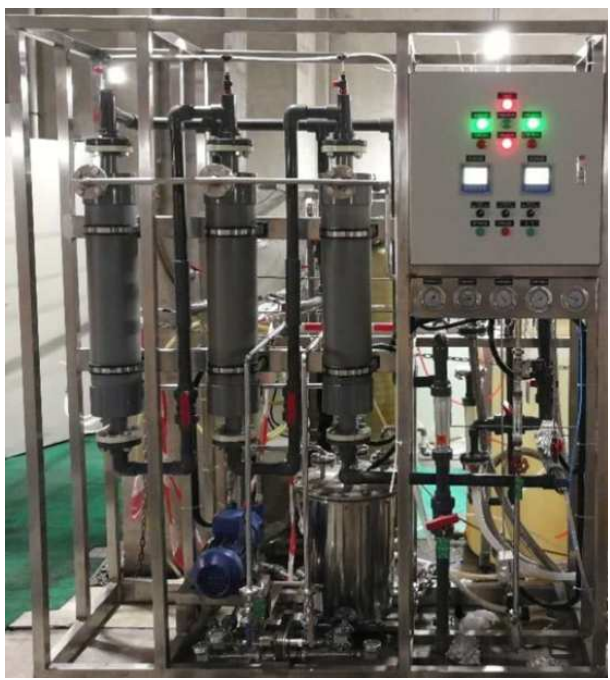}
\caption{The real picture of the degasser membrane.}
\label{fig.realdegasser}
\end{figure}

A prototype detector has been built to test the key technical issues as well as to test the performance of the water system of JUNO. Fig.~\ref{fig.realdegasser} shows the real picture of the degassing membranes, which have been installed in JUNO prototype.

\section{The radon measurement system}

\begin{figure}[htb]
\centering
\includegraphics[width=8cm]{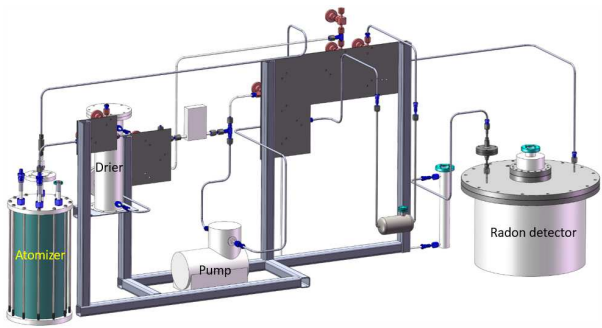}
\caption{The conceptual diagram of the radon measurement system.}
\label{fig.radonmeasurement}
\end{figure}

A  high sensitivity radon concentration measurement system has been developed to measure the radon concentration in water. The conceptual diagram is shown in Fig.~\ref{fig.radonmeasurement} and the detail information about the detector can be found in Ref.~\cite{RDTM}. The atomizer and radon detector are the two most important parts of the system.

\begin{figure}[htb]
\centering
\includegraphics[width=6cm]{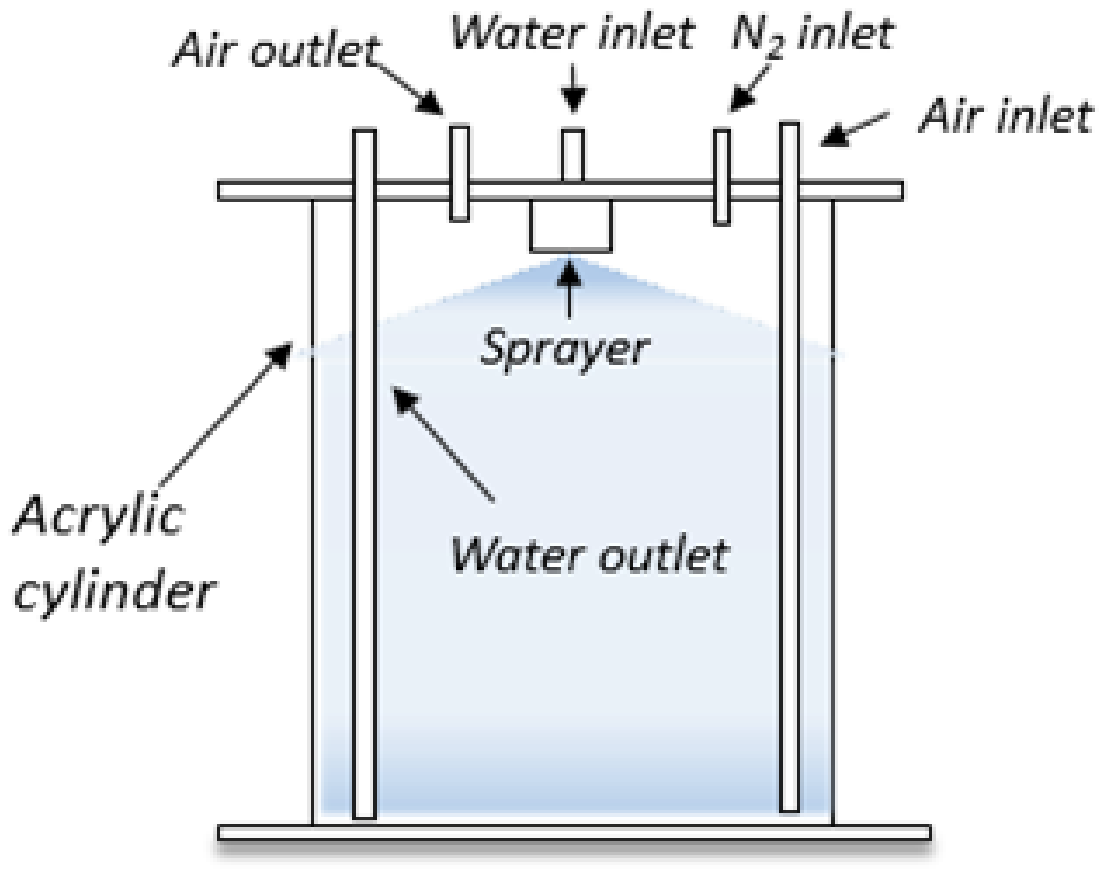}
\includegraphics[width=6cm]{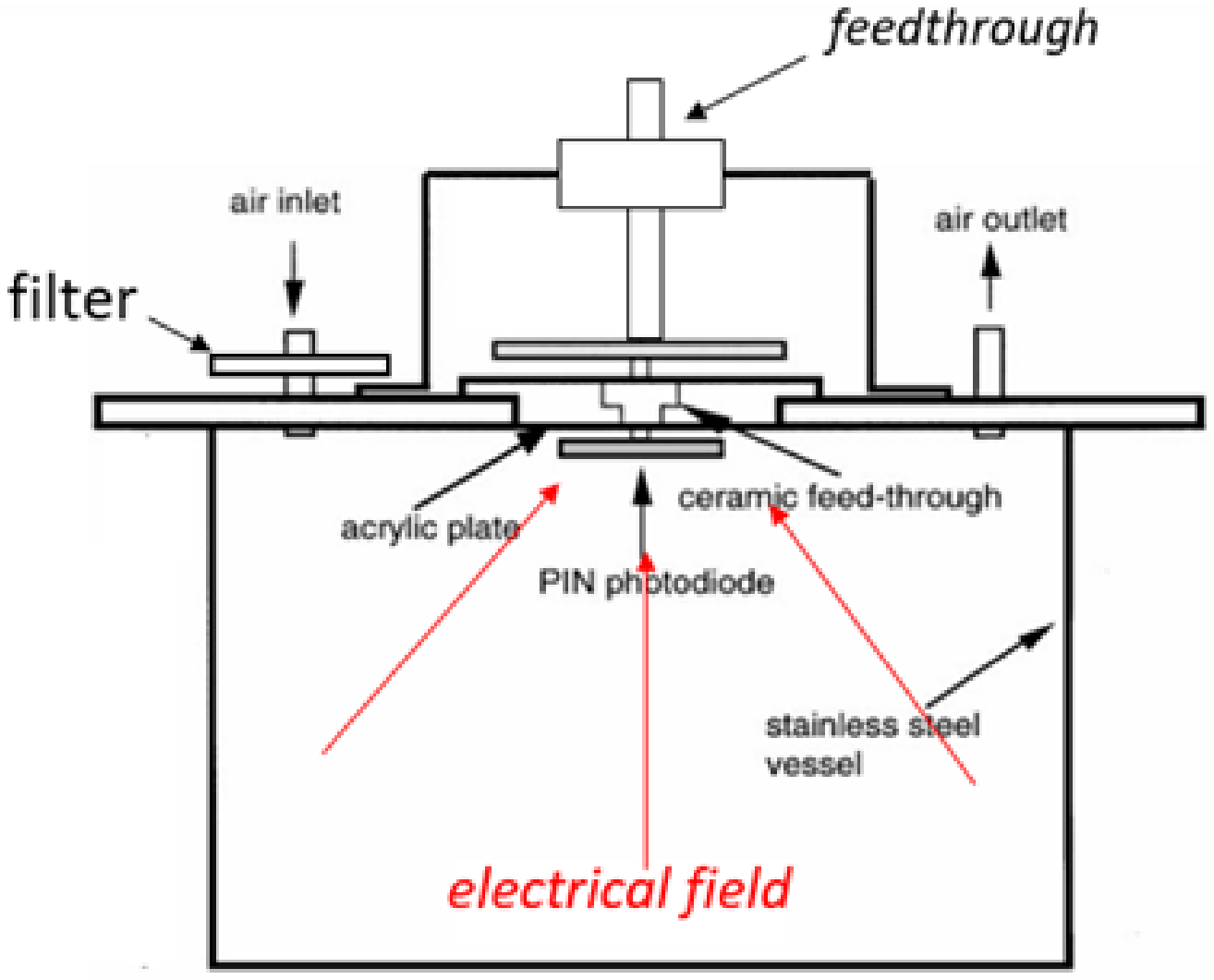}
\caption{Left: The schematic diagram of the atomizer. Right: The schematic diagram of the radon detector.}
\label{fig.energy response}
\end{figure}

(A) The atomizer is a water vapor balancing device and it can be used to transfer the radon from water into vapor by spraying. When the water and vapor are at equilibrium state, the radon concentration in the two media(water/vapor) is correlated by the Ostwald coefficient and the ratio of radon concentrations in water to vapor follows Equ.~\ref{equ.Rn_ratio}~\cite{Rn_ratio_rad7}:

\begin{equation}
R = 0.105 + 0.405e^{-0.0502T}
\label{equ.Rn_ratio}
\end{equation}

where R stands for the ratio and T is the temperature in unit of centigrade. For JUNO prototype, the temperature of the water has been kept at around 20 centigrade; thus, R is around 0.25. Therefore, the Rn concentration in water can be derived from the gas measurement result.

(B)The radon detector consists of a cylindrical electro-polished stainless steel vessel, a cylindrical high-purity oxygen-free copper vessel, and a Si-PIN photodiode. The principle of the radon detector is to collect the daughter nuclei of $^{222}$Rn to the surface of the Si-PIN with an electric field, and to measure the energy of $\alpha$s released by the collected nuclei~\cite{annreport,radioisotopes}. The sensitivity of the radon detector is around 9.0~mBq/m$^{3}$ for a single-day measurement~\cite{RDTM}.

(C) The drier is used to keep the relative humidity below 3\% because 90\% of the radon daughters are positive and the high humidity will decrease the collecting efficiency~\cite{NIMA_421}. The Pump is used to circulate the gas in the measurement system.

\section{Measurement results}

\begin{figure}[htb]
\centering
\includegraphics[width=8cm]{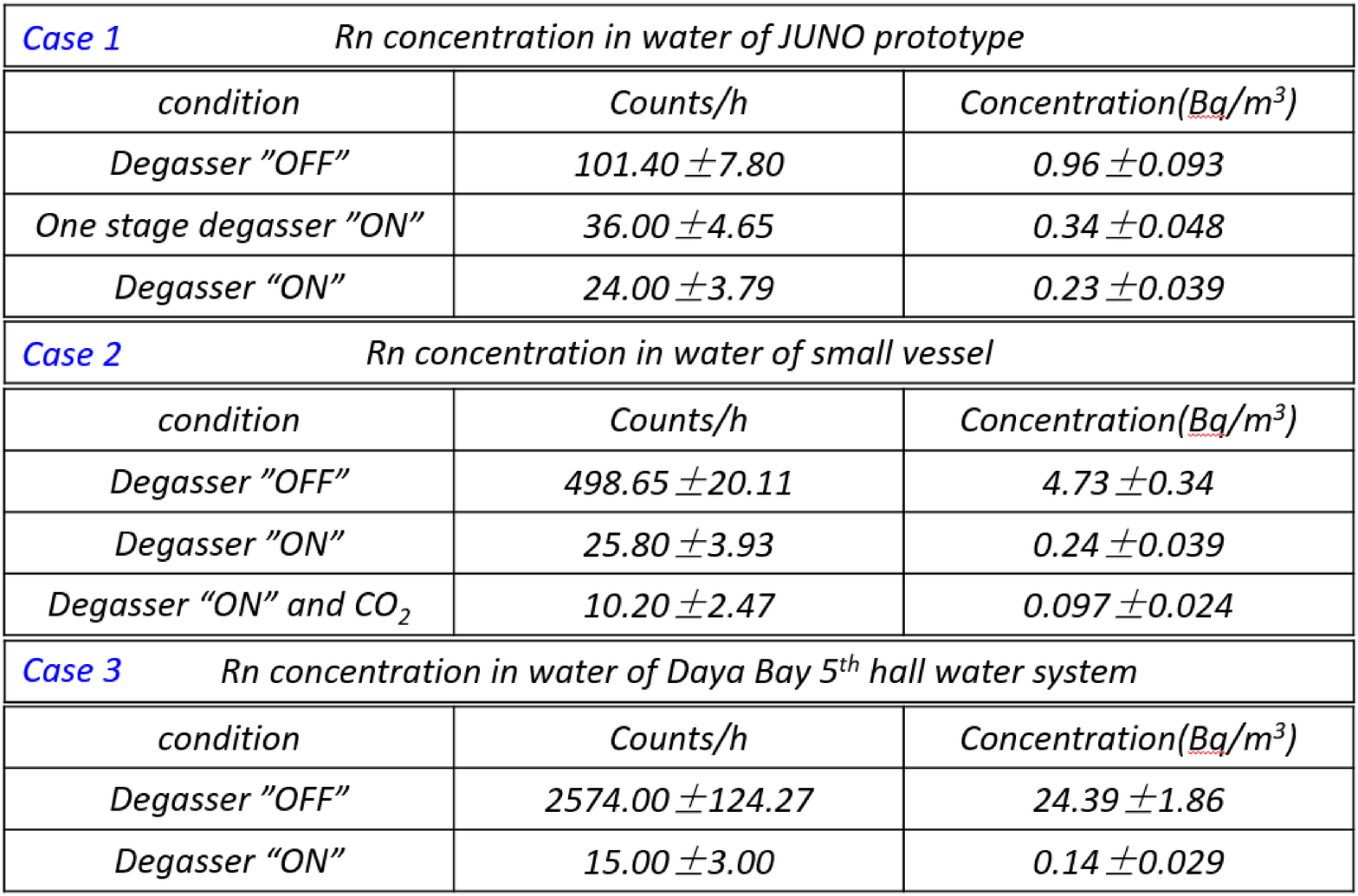}
\caption{The measurement results of radon concentration in water at different conditions. Case 1: 0.15MPa inlet water pressure; Case 2: 0.15MPa inlet water pressure with carbon dioxide loaded; Case 3: 0.35MPa inlet water pressure.}
\label{fig.results}
\end{figure}

The radon removal efficiency has been tested at different conditions. In case 1, the degassing membranes are tested with the JUNO prototype water system at 0.15MPa inlet water pressure and $\sim$0.5m$^{3}$/h flow rate, the results show that the degassing efficiency for one stage of the degassing membrane is around 65\%. While after several times of degassing, the radon concentration is still around 0.23~Bq/m$^{3}$, which implies that the radon removal limit of the degassing membranes can not satisfy the JUNO requirement at this condition.

In order to improve the radon removal efficiency as well as lower the radon removal limit, two kinds of different methods have been put forward, namely loading carbon dioxide into the water and increasing the inlet water pressure. The carbon dioxide loading flow rate is $\sim$0.3~L/min and the water flow rate for the two cases are $\sim$0.5m$^{3}/h$. The results show that both of the two kinds of methods could help to lower the radon removal limit and the radon concentration in water can be reduced to around 0.1Bq/m$^{3}$, which can satisfy the requirement of JUNO.

\section{Summary}
The main goal of JUNO is to determine the neutrino mass hierarchy and the cosmic ray muons are the source of the main background. The water Cherenkov detector can be used as passive shielding and muon veto, thus a reliable ultra-pure water production and circulation system has been designed to keep the water quality good and to keep the radon concentration below 0.2Bq/m$^{3}$. A high sensitivity radon detector has been developed for radon concentration measurement and the Liqui-Cel degassing membranes are used to remove the radon in water. Loading carbon dioxide into the water and increasing the inlet water pressure could help to lower the radon removal limit of the degassing membrane and the radon concentration in water can be reduced to around 0.1Bq/m$^{3}$, which can satisfy the requirement of JUNO.

\section{Acknowledgements}
Many thanks to Shoukang Qiu and Quan Tang of University of South China for their help during the experiment.

\end{document}